\documentclass[aps,prl,reprint,superscriptaddress,showpacs]{revtex4-1}

\usepackage{graphicx}
\usepackage{dcolumn}
\usepackage{bm}
\usepackage{slashed}
\usepackage{hyperref}
\usepackage[mathlines]{lineno}

\begin{document}


\title{First measurement of proton's charge form factor at very low $Q^2$ with initial state radiation}



\author{M.~Mihovilovi\v{c}}
  \affiliation{Institut~f\"{u}r~Kernphysik, Johannes~Gutenberg-Universit\"{a}t~Mainz, DE-55128~Mainz,~Germany}
  \affiliation{Jo\v{z}ef~Stefan~Institute, SI-1000 Ljubljana, Slovenia}

\author{A.~B.~Weber}
  \affiliation{Institut~f\"{u}r~Kernphysik, Johannes~Gutenberg-Universit\"{a}t~Mainz, DE-55128~Mainz,~Germany}

\author{P.~Achenbach}
  \affiliation{Institut~f\"{u}r~Kernphysik, Johannes~Gutenberg-Universit\"{a}t~Mainz, DE-55128~Mainz,~Germany}
  
\author{T.~Beranek}
  \affiliation{Institut~f\"{u}r~Kernphysik, Johannes~Gutenberg-Universit\"{a}t~Mainz, DE-55128~Mainz,~Germany}
  
\author{J.~Beri\v{c}i\v{c}}
  \affiliation{Jo\v{z}ef~Stefan~Institute, SI-1000 Ljubljana, Slovenia}
  
\author{J.~C.~Bernauer}
  \affiliation{Massachusetts Institute of Technology, Cambridge, MA~02139, USA}
  
\author{R.~B\"{o}hm}
  \affiliation{Institut~f\"{u}r~Kernphysik, Johannes~Gutenberg-Universit\"{a}t~Mainz, DE-55128~Mainz,~Germany}
  
\author{D.~Bosnar}
 \affiliation{Department~of~Physics, University~of~Zagreb, HR-10002~Zagreb, Croatia}
  
\author{M.~Cardinali}
  \affiliation{Institut~f\"{u}r~Kernphysik, Johannes~Gutenberg-Universit\"{a}t~Mainz, DE-55128~Mainz,~Germany}
  
\author{L.~Correa}
 \affiliation{Universit\'{e}~Clermont~Auvergne, CNRS/IN2P3, LPC, BP~10448, F-63000 Clermont-Ferrand, France}
 
\author{L.~Debenjak}
  \affiliation{Jo\v{z}ef~Stefan~Institute, SI-1000 Ljubljana, Slovenia}

\author{A.~Denig}
  \affiliation{Institut~f\"{u}r~Kernphysik, Johannes~Gutenberg-Universit\"{a}t~Mainz, DE-55128~Mainz,~Germany}
  
\author{M.~O.~Distler}
  \affiliation{Institut~f\"{u}r~Kernphysik, Johannes~Gutenberg-Universit\"{a}t~Mainz, DE-55128~Mainz,~Germany}
  
\author{A.~Esser}
  \affiliation{Institut~f\"{u}r~Kernphysik, Johannes~Gutenberg-Universit\"{a}t~Mainz, DE-55128~Mainz,~Germany}
  
\author{M.~I.~Ferretti~Bondy}
  \affiliation{Institut~f\"{u}r~Kernphysik, Johannes~Gutenberg-Universit\"{a}t~Mainz, DE-55128~Mainz,~Germany}
  
\author{H.~Fonvieille}
 \affiliation{Universit\'{e}~Clermont~Auvergne, CNRS/IN2P3, LPC, BP~10448, F-63000 Clermont-Ferrand, France}

\author{J.~M.~Friedrich}
  \affiliation{Technische Universit\"{a}t M\"{u}nchen, Physik Department, 85748 Garching, Germany}

\author{I.~Fri\v{s}\v{c}i\'{c}}
 \affiliation{Department~of~Physics, University~of~Zagreb, HR-10002~Zagreb, Croatia}

\author{K.~Griffioen}
 \affiliation{College of William and Mary, Williamsburg, VA~23187, USA}

\author{M.~Hoek}
  \affiliation{Institut~f\"{u}r~Kernphysik, Johannes~Gutenberg-Universit\"{a}t~Mainz, DE-55128~Mainz,~Germany}
  
\author{S.~Kegel}
  \affiliation{Institut~f\"{u}r~Kernphysik, Johannes~Gutenberg-Universit\"{a}t~Mainz, DE-55128~Mainz,~Germany}
  
\author{Y.~Kohl}
  \affiliation{Institut~f\"{u}r~Kernphysik, Johannes~Gutenberg-Universit\"{a}t~Mainz, DE-55128~Mainz,~Germany}
  
\author{H.~Merkel}
  \email{merkel@kph.uni-mainz.de}
  \affiliation{Institut~f\"{u}r~Kernphysik, Johannes~Gutenberg-Universit\"{a}t~Mainz, DE-55128~Mainz,~Germany}
  
\author{D.~G.~Middleton}
  \affiliation{Institut~f\"{u}r~Kernphysik, Johannes~Gutenberg-Universit\"{a}t~Mainz, DE-55128~Mainz,~Germany}
      
\author{U.~M\"{u}ller}
  \affiliation{Institut~f\"{u}r~Kernphysik, Johannes~Gutenberg-Universit\"{a}t~Mainz, DE-55128~Mainz,~Germany}

\author{L.~Nungesser}
  \affiliation{Institut~f\"{u}r~Kernphysik, Johannes~Gutenberg-Universit\"{a}t~Mainz, DE-55128~Mainz,~Germany}
  
\author{J.~Pochodzalla}
  \affiliation{Institut~f\"{u}r~Kernphysik, Johannes~Gutenberg-Universit\"{a}t~Mainz, DE-55128~Mainz,~Germany}
  
\author{M.~Rohrbeck}
  \affiliation{Institut~f\"{u}r~Kernphysik, Johannes~Gutenberg-Universit\"{a}t~Mainz, DE-55128~Mainz,~Germany}
  
\author{S.~S\'anchez~Majos}
  \affiliation{Institut~f\"{u}r~Kernphysik, Johannes~Gutenberg-Universit\"{a}t~Mainz, DE-55128~Mainz,~Germany}  
  
\author{B.~S.~Schlimme}
  \affiliation{Institut~f\"{u}r~Kernphysik, Johannes~Gutenberg-Universit\"{a}t~Mainz, DE-55128~Mainz,~Germany}
  
\author{M.~Schoth}
  \affiliation{Institut~f\"{u}r~Kernphysik, Johannes~Gutenberg-Universit\"{a}t~Mainz, DE-55128~Mainz,~Germany}
  
\author{F.~Schulz}
  \affiliation{Institut~f\"{u}r~Kernphysik, Johannes~Gutenberg-Universit\"{a}t~Mainz, DE-55128~Mainz,~Germany}
  
\author{C.~Sfienti}
  \affiliation{Institut~f\"{u}r~Kernphysik, Johannes~Gutenberg-Universit\"{a}t~Mainz, DE-55128~Mainz,~Germany}
  
\author{S.~\v{S}irca}
  \affiliation{Faculty~of~Mathematics~and~Physics, University~of~Ljubljana, SI-1000 Ljubljana, Slovenia}
  \affiliation{Jo\v{z}ef~Stefan~Institute, SI-1000 Ljubljana, Slovenia}

\author{S.~\v{S}tajner}
  \affiliation{Jo\v{z}ef~Stefan~Institute, SI-1000 Ljubljana, Slovenia}

\author{M.~Thiel}
  \affiliation{Institut~f\"{u}r~Kernphysik, Johannes~Gutenberg-Universit\"{a}t~Mainz, DE-55128~Mainz,~Germany}
  
\author{A.~Tyukin}
  \affiliation{Institut~f\"{u}r~Kernphysik, Johannes~Gutenberg-Universit\"{a}t~Mainz, DE-55128~Mainz,~Germany}
  
\author{M.~Vanderhaeghen}
  \affiliation{Institut~f\"{u}r~Kernphysik, Johannes~Gutenberg-Universit\"{a}t~Mainz, DE-55128~Mainz,~Germany}
  
\author{M.~Weinriefer}
  \affiliation{Institut~f\"{u}r~Kernphysik, Johannes~Gutenberg-Universit\"{a}t~Mainz, DE-55128~Mainz,~Germany}
\collaboration{A1-Collaboration}\noaffiliation

\date{\today}

\begin{abstract}
We report on a new experimental method based on 
initial-state radiation (ISR) in $e$-$p$ scattering, in which the radiative tail of the 
elastic $e$-$p$ peak contains information on the proton charge form factor ($G_E^{p}$) 
at extremely small $Q^2$. The ISR technique was validated in a dedicated experiment 
using the spectrometers of the A1-Collaboration at the Mainz Microtron (MAMI). This 
provided first measurements of $G_E^{p}$ for $0.001\leq Q^2 \leq 0.004\,(\mathrm{GeV}/c)^2$.
\end{abstract}


{\boldmath \maketitle}

\section{Introduction}
The radius of the proton as a fundamental subatomic constant has recently 
received immense attention. 
The CODATA~\cite{CODATA2014} value of $0.8751(61)\,\mathrm{fm} $
was compiled from electron scattering and atomic Lamb shift measurements. 
Both approaches gave consistent results.  This value however, does not agree with the 
findings of very precise Lamb shift measurements in muonic 
hydrogen~\cite{Pohl2010, Antognini2013},  which are $6\,\sigma$ away 
from the CODATA value. This discrepancy cannot be explained within 
existing physics theories, nor can it be interpreted as an experimental error.
To provide further insight into the matter, several new spectroscopic and 
scattering experiments are underway. They aim to investigate different aspects 
of the problem~\cite{PohlAnn,Carlson201559}.

In a scattering experiment the charge radius of the proton is typically determined by 
measuring cross sections for elastic scattering of electrons from hydrogen, which 
depend on $G_E^p$ and carry information about the charge 
distribution in the proton. The proton charge radius is given by
\begin{eqnarray}
r_p^2 \equiv \left.-6\hbar^2 \frac{\mathrm{d}G_E^p}{\mathrm{d}Q^2}\right |_{Q^2=0}\,,\label{pr1}
\end{eqnarray}
where $Q^2$ is the negative square of the four-momentum transferred to the hadron. 
 Due to the limited reach of existing data sets
($Q^2 > 0.004\,\mathrm{GeV}^2/c^2$) the slope of $G_E^p$ at $Q^2=0$
needs to be evaluated from an extrapolated fit of the measured data. 
The available data have enough resolving power to precisely determine 
the slope of the form factor at some distance from the origin, but additional 
data are needed to constrain the slope at $Q^2=0$. Therefore, measurements of $G_E^p$
need to be extended into the previously unmeasured region of  
$Q^2\lesssim 10^{-3}\,\mathrm{GeV}^2/c^2$.

Efforts to do such measurements with the standard approaches are
limited by the minimal $Q^2$ accessible with the experimental apparatus
at hand. The energy of the electron beam and the scattering angle must 
be very small. Here we present a new experimental approach
that avoids these kinematic limitations, extends the currently accessible $Q^2$ range, 
and allows for cross section measurements below $0.004\,\mathrm{GeV^2}/c^2$ with 
sub-percent precision. The  initial state radiation (ISR) technique 
exploits information within the radiative tail of the elastic peak. 
This was inspired by a similar concept used in particle physics to measure 
$e^{+}e^{-}\rightarrow \mathrm{hadrons}$ over a wide range of center-of-mass 
energies in a single experiment~\cite{Arbuzov1998,BabarJPSi}.

\section{Initial state radiation Technique}

\begin{figure}[ht]
 \includegraphics[height=0.35\textwidth]{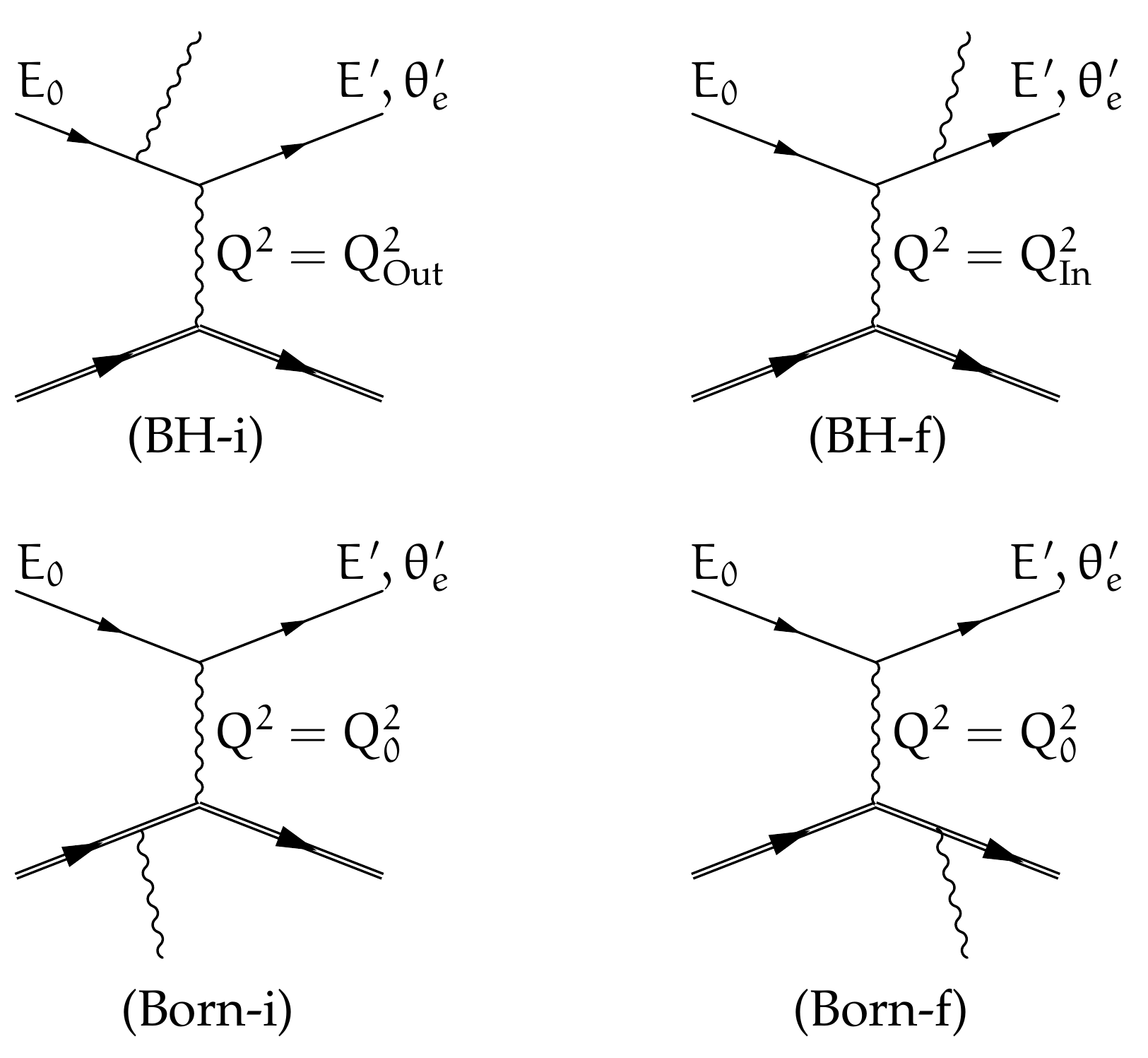}
 \caption{Feynman diagrams for inelastic scattering of an electron from a proton, 
 where an electron or a proton emits a real photon before or after the 
 interaction. Diagrams where electrons emit a photon are known as Bethe-Heitler 
 (BH) diagrams, while those where protons emit real photons are called Born 
 diagrams. $Q^2_{\mathrm{In}}$ is the squared four-momentum fixed by the 
 beam energy and the scattering angle, while $Q^2_{\mathrm{Out}}\leq Q^2_{\mathrm{In}}$
 corresponds to the value measured with the detector. $Q_0^2 = 4E_0 E' \sin^2{(\theta_e'/2)}/c^2$,
 where $E_0$ is the energy of the incident electron, $E'$ and $\theta_e'$ are the energy and 
 angle of the detected electron. For the (BH-i) diagram $Q^2 = Q^2_{\mathrm{Out}}$, 
 and for the (BH-f) diagram $Q^2 = Q^2_{\mathrm{In}}$.
 \label{fig_ISRDiagrams}}
\end{figure}

The radiative tail of an elastic peak is dominated by the contributions from two Bethe-Heitler
diagrams~\cite{Vanderhaeghen2000} as shown in Fig.~\ref{fig_ISRDiagrams}. 
The initial-state radiation (BH-i) corresponds to 
the incident electron emitting a real photon before interacting with the proton,
and the final-state radiation  (BH-f) corresponds to a real photon being 
emitted after the interaction with the nucleon. For these processes two characteristic 
$Q^2$ can be defined:
\begin{eqnarray}
Q^2_{\mathrm{In}} = \frac{4\frac{E_0^2}{c^2}\sin^2{\frac{\theta_{e}'}{2}}}{1+\frac{2E_0}{Mc^2}\sin^2{\frac{\theta_{e}'}{2}}}\,\>\>\mathrm{and}\>\>
Q^2_{\mathrm{Out}} = \frac{4\frac{E'^2}{c^2}\sin^2{\frac{\theta_{e}'}{2}}}{1-\frac{2E'}{Mc^2}\sin^2{\frac{\theta_{e}'}{2}}}\,. \nonumber
\end{eqnarray}
Here, $Q^2_{\mathrm{In}}$ represents the value set by the chosen kinematics for elastic scattering 
($E_0$, $\theta_e'$), while $Q^2_{\mathrm{Out}}$ corresponds to the value measured by 
the detectors after scattering. $E_0$ and $E'$ are the energies of the incoming and scattered electrons, $M$ is the mass of the proton, and $\theta_{e}'$ is the scattering angle 
of the detected electron. In the limit of exact elastic $\mathrm{H}(e,e')p$ scattering, 
$Q^2_{\mathrm{In}}$ and $Q^2_{\mathrm{Out}}$ are both equal to $Q^2_0 = 4E_0 E' \sin^2{(\theta_e'/2)}/c^2$ and correspond to the $Q^2$ actually transferred to the proton. 
In $\mathrm{H}(e,e')\gamma p$, however,  $Q^2_{\mathrm{In}}$ and $Q^2_{\mathrm{Out}}$ no longer coincide. 
In the initial-state radiation diagram the emitted photon 
carries away part of the incident electron's four-momentum and 
opens the possibility to probe the proton's electromagnetic 
structure at $Q^2 = Q^2_{\mathrm{Out}}$ which is smaller than
$Q^2_{\mathrm{In}}$. On the other hand, in the final-state radiation diagram the 
momentum transfer at the vertex remains fixed ($Q^2 = Q^2_{\mathrm{In}}$),
thus only $Q^2_{\mathrm{Out}}$ is modified, $Q^2_{\mathrm{Out}}\leq Q^2$. 

In an inclusive $(e,e')$ experiment only $Q^2_{\mathrm{Out}}$ can be measured, 
which implies that initial state radiation cannot 
be distinguished from final state radiation. The measured radiative 
tail represents an approximately  $2:3$ mixture of terms with 
$Q^2=Q^2_{\mathrm{In}}$ and $Q^2=Q^2_{\mathrm{Out}}$, respectively.  
There are also Born terms (Born-i and Born-f), where 
the initial and final protons emit real photons, as well as higher-order  
radiative corrections that also contribute to the radiative tail.
The basic concept of the ISR approach is to isolate the interesting 
(BH-i) process from other contributions to the radiative tail, and thus 
obtain information on form factors at unmeasured values of $Q^2=Q^2_{\mathrm{Out}}$. To accomplish this,
the measurements need to be studied in conjunction with a Monte-Carlo simulation
that encompasses a comprehensive description of the radiative tail. 

\section{Description of the Radiative tail}

To realistically mimic the radiative tail, the peaking approximation models
devised from the corrections to the elastic cross section are insufficient~\cite{Vanderhaeghen2000}.  
For an adequate description far away from the elastic line 
($Q^2=Q^2_{\mathrm{Out}}\ll Q^2_{\mathrm{In}}$), it is crucial to consider 
cross-section contributions
to the $e^8$-order. To achieve this goal, a Monte-Carlo simulation is used,  
which employs a sophisticated event generator that calculates 
amplitudes  exactly for the leading, $e^3$-order 
diagrams (shown in Fig.~\ref{fig_ISRDiagrams}) and includes $G_E^p$ as a free, tuneable 
parameter for every simulated $Q^2$.  
The next order vacuum polarization diagrams (with electrons inside the fermion loop) 
are exactly calculable and are added as a multiplicative factor to the cross section. 
The virtual corrections to the Bethe-Heitler diagrams (self-energy corrections and 
various vertex corrections) require integration of the loop diagrams and are 
computationally too intensive to be added directly to the simulation. Instead 
they are considered as effective corrections to the cross section using 
the prescription of Ref.~\cite{Vanderhaeghen2000}, together with the real second-order 
correction (emission of two real photons) which is approximated using the corrections to the 
elastic cross section~\cite{Vanderhaeghen2000, Maximon2000}. 
Hadronic corrections are also considered in the elastic limit
using the calculations of Ref.~\cite{Maximon2000}. They contribute only up 
to $0.5\,\mathrm{\%}$ to the cross section at the lowest energy settings. In the 
simulation the proton is always on-shell. Effects related to the 
internal structure of the proton, described by the general 
polarisabilities~\cite{ARENHOVEL} and known from the virtual 
Compton scattering (VCS) experiments~\cite{Roche2000}, were 
small and could be neglected. Besides 
the internal corrections, the simulation includes 
external radiative and Coulomb corrections~\cite{MoTsai, Tsai}, 
collisional losses of particles on their way from the vertex point 
to the detectors, and the precise acceptances of the spectrometers.   


\section{Experiment}

The measurement of 
the radiative tail has been performed at the Mainz Microtron (MAMI) in 2013 using 
the spectrometer setup of the A1-Collaboration~\cite{Blomqvist}. In the 
experiment a rastered electron beam with energies of $195$, 
$330$ and $495\,\mathrm{MeV}$ was used in combination with a 
hydrogen target, which consisted of a $5\,\mathrm{cm}$-long cigar-shaped 
Havar cell filled with liquid hydrogen and
placed in an evacuated scattering chamber. For the cross section measurements 
the single-dipole magnetic spectrometer~B with a momentum
acceptance of $\pm 7.5\,\mathrm{\%}$ was employed. It was positioned at a fixed 
angle of $15.21^\circ$, while its momentum settings were adjusted to scan 
the complete radiative tail for each beam energy. The central momentum of each 
setting was measured with an NMR probe to a relative accuracy of $8\times10^{-5}$. 
The spectrometer is equipped with a detector package consisting of two layers of 
vertical drift chambers (VDCs) for tracking, two layers of scintillation detectors 
for triggering, and a threshold Cherenkov detector for particle identification. 
The kinematic settings of the experiment were chosen such that the radiative 
tails recorded at three beam energies overlap. 

The beam current was between $10\,\mathrm{nA}$ and $1\,\mu\mathrm{A}$
and was limited by the maximum rate allowed  in the VDCs 
($\approx 1\,\mathrm{kHz/wire}$), resulting in raw rates 
up to $20\,\mathrm{kHz}$. The current was determined by 
a non-invasive fluxgate-magnetometer and from the collected charge of the stopped 
beam. At low beam currents and low beam energies the accuracy of
both approaches is not better than $ 2\,\mathrm{\%}$, which is insufficient for 
a precision cross section measurement. Hence spectrometer~A, used in a fixed 
momentum and angular setting, was employed for precise monitoring of the 
relative luminosity.

In spite of the good vacuum conditions inside the scattering chamber 
($10^{-6}\,\mathrm{mbar}$), the experiment was sensitive to traces of 
cryogenic depositions on the target walls, 
consisting mostly of residual nitrogen and oxygen present in the scattering 
chamber~\cite{MihovilovicMessina}. Since the deposited layer affected the 
measured spectra, the kinematic settings for spectrometer A were chosen such 
that the nitrogen/oxygen elastic lines were always visible next to the hydrogen 
spectrum, which served as a precise monitor of the thickness of the cryogenic 
depositions.

The data were collected at a rate of $800$ events per second and with a live-time of 
 about $50\,\mathrm{\%}$. Each collected data sample contains 
about $2\,\mathrm{M}$ events and consists of measurements of the radiative 
tail for a chosen $E'$ range collected with spectrometer B and a corresponding 
reference (luminosity) spectrum from spectrometer A.

\section{Data Analysis}

Measurements at the highest beam energy settings encompass the range of $Q^2$
where $G_E^{p}$ is known from previous experiments, and were then used for the validation 
of the ISR technique. The measurements with the beam energies of $330\,\mathrm{MeV}$ and $195\,\mathrm{MeV}$ were used to investigate $G_E^{p}$ at previously unattained values of $Q^2$.
 

Before comparing the data to the simulation, the measured spectra had to be corrected
for the inefficiencies of the detection system. The efficiencies of the scintillation 
detector and the Cherenkov detector were determined to be $(99.8\pm0.2)\,\mathrm{\%}$
and  $(99.74\pm0.02)\,\mathrm{\%}$, respectively, and were considered as multiplicative
correction factors to the measured distributions. The quality of the agreement 
between the data and simulation depends also on the momentum and 
spatial resolutions of the spectrometer. These were determined 
from dedicated calibration data sets. The relative momentum plus angular and vertex  
resolutions (FWHM) were $1.7\times 10^{-4}$, $3\,\mathrm{msr}$, and $1.6\,\mathrm{mm}$,
respectively.

A series of cuts were applied to the data in order to minimize the background.
First, a cut on the Cherenkov signal was applied to identify electrons,
followed by a cut on the nominal momentum acceptance of the spectrometer.  
To minimize the contributions of events coming from the target walls and 
cryogenic depositions, a rather strict, $\pm 10\,\mathrm{mm}$ cut on the vertex 
position was applied. Due to the finite vertex resolution some of the background 
events remained in the cut sample. Their contribution to the spectra was 
estimated by using a dedicated simulation, normalized to the size 
of the nitrogen, oxygen and Havar elastic lines, and corrected for the changes 
in the thickness of the depositions versus time by using the data of Spectrometer A.

The most challenging background came from the 
entrance flange of spectrometer B and the metal support structure of the 
target cell.  When measuring far away from the elastic peak, the elastically scattered electrons, 
which a priori are not accepted, undergo secondary processes in these 
components and re-scatter into the acceptance of the spectrometer. At
high $E'$ these contributions are negligible, but at low $E'$, where the cross
section for the Bethe-Heitler processes becomes comparable to the probability
for double scattering, these secondary reactions begin to contribute 
substantially to the detected number of events.  At high beam energy 
settings, the background can be successfully removed via strict cuts on vertex and 
out-of-plane angle. However, at the lowest energy settings, a substantial part 
remained inside the data, which limited our efforts to measure at lower $Q^2$.
Since this background could not be adequately subtracted or simulated, the 
data with $E'< 128\,\mathrm{MeV}$ were omitted from the present analysis,
which limited the reach of the experiment to $Q^2 \geq 1.3\cdot 10^{-3}\,\mathrm{GeV}^2/c^2$.


Additionally, the external radiative corrections are not considered
to the same order of precision as the internal radiative corrections.
This is not problematic in the region of the tail, where the size
of the former is small.  However, in the immediate vicinity of
the elastic peak, where their contribution is substantial, they
may result in an incorrect description of the momentum distribution.
To avoid this problem, the unradiated elastic data (from the first bin)
were omitted from the analysis.  

\begin{figure}[ht!]
 \includegraphics[width=0.48\textwidth]{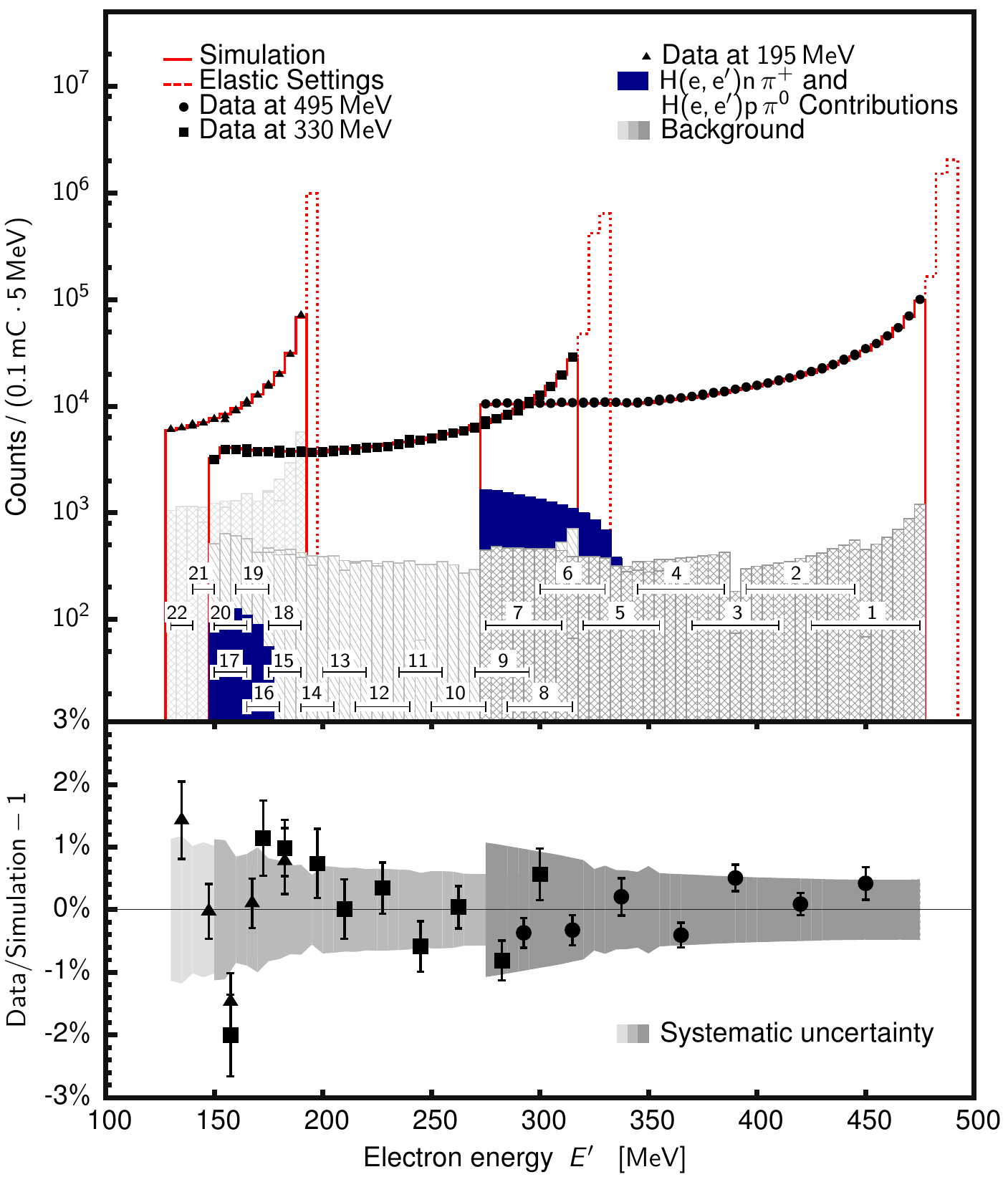}
 \caption{(Color on-line) Comparison of the data to the simulation. 
 {\bf Top:} Circles, squares and triangles show the measured 
 distributions at $495\,\mathrm{MeV}$,  $330\,\mathrm{MeV}$ and 
 $195\,\mathrm{MeV}$, respectively, normalized to the 
 accumulated charge of $0.1\,\mathrm{mC}$. 
 The elastic data (dashed line) are omitted from the analysis. 
 The simulations with $G_E^p$, given by  parameterization of Bernauer~\cite{Bernauer2014} 
 are shown with red lines.  
 The measurements at $495\,\mathrm{MeV}$, $330\,\mathrm{MeV}$, $195\,\mathrm{MeV}$ 
 were divided into seven ($1-7$), ten ($8-17$) and five ($18-22$) energy bins, 
 respectively, such that two neighboring settings overlap for a half of the energy acceptance. 
 The residual contributions of target walls, target frame, spectrometer entrance flange 
 and cryogenic depositions are shown with shaded areas. The full (blue) areas represent the 
 contributions of the pion production processes. {\bf Bottom:} Relative difference between 
 the data and simulation. The points show the mean values for each kinematic point, while 
 the error bars denote their statistical uncertainties. Gray bands demonstrate the systematic uncertainties.  
 \label{fig_Results}}
\end{figure}

The cleaned event samples for each kinematic setting were  
corrected for the dead-time and prescale factors, weighted by the relative 
luminosity determined by spectrometer A, and then merged together to form 
a single spectrum that could be compared to the simulation (see Fig.~\ref{fig_Results}).
The simulation was performed with the Bernauer parameterization of  
$G_{E}^{p}$~\cite{Bernauer2014}. The contribution of $G_{M}^{p}$  to the cross section 
at $Q^2\leq 10^{-2}\,\mathrm{GeV^2}/c^2$ is smaller than 
$0.5\,\mathrm{\%}$ and can therefore be approximated with the standard dipole 
model.
For each beam-energy setting 
golden data were selected which served as a reference for the relative normalization 
of luminosity for other data sets. Hence, for each of the three beam energies one  
parameter (absolute luminosity) remained unknown and was fixed by equating the average 
ratio of data to simulation to unity. 

In the bins far away from the elastic peak, one also needs to consider 
$\mathrm{H}(e,e')n\pi^{+}$ and $\mathrm{H}(e,e')p\pi^{0}$ reactions, which contribute 
up to $10\,\mathrm{\%}$ of all events. These processes were simulated using the MAID 
model~\cite{MAID} and were added to the full simulation before comparing 
it to the data.

\section{Systematic uncertainties}
The ISR technique provides remarkable control over the
systematic uncertainties. With the fixed angular settings and overlapping 
momentum ranges all ambiguities related to the acceptances disappear. 
Furthermore, the luminosity is directly measured with spectrometer A,
thus avoiding potential problems with fluctuations in the beam current and 
target density.  The relative luminosity is determined 
with an accuracy better than $0.17\,\mathrm{\%}$.  
Other sources of systematic uncertainty are: the ambiguity in the 
determination of detector efficiencies $(0.2\,\mathrm{\%})$;
the inconclusiveness of the background simulation at lowest momenta 
$(\leq 0.5\,\mathrm{\%})$; the contribution of higher-order 
corrections, which are not included in the simulation $(0.3\,\mathrm{\%})$; 
and the contamination with events coming from the target support frame and 
the spectrometer entrance flange $(0.4\,\mathrm{\%})$.  
The bins containing pions are subjected to another $0.5\,\mathrm{\%}$ 
uncertainty of the MAID model near the pion production threshold. 
This contribution, which appears to be an 
important source of the systematic uncertainty, is significant only for the
$495\,\mathrm{MeV}$ setting. For the measurements at $195\,\mathrm{MeV}$
and $330\,\mathrm{MeV}$ the contribution of pion production 
processes is less than $2\,\mathrm{\%}$ and the corresponding systematic uncertainty 
is $\leq 0.1\,\mathrm{\%}$.

\section{Results and Outlook}
The ratio of measured and calculated cross sections shown in Fig.~\ref{fig_Results} (bottom) 
are in agreement to within a percent for all three energies. Considering the Bernauer fit~\cite{Bernauer2014} 
as a credible description of $G_E^p$ this demonstrates for the first time that the 
electromagnetic processes, which give rise to the radiative tail are understood to a few 
parts per thousand, even at  $200\,\mathrm{MeV}$ below the elastic line. This is an important 
finding for the electron-induced experiments, such as VCS~\cite{Janssens2008}, which 
require precise knowledge of the radiative corrections. 

\begin{figure*}[ht!]
 \includegraphics[width=0.8\textwidth]{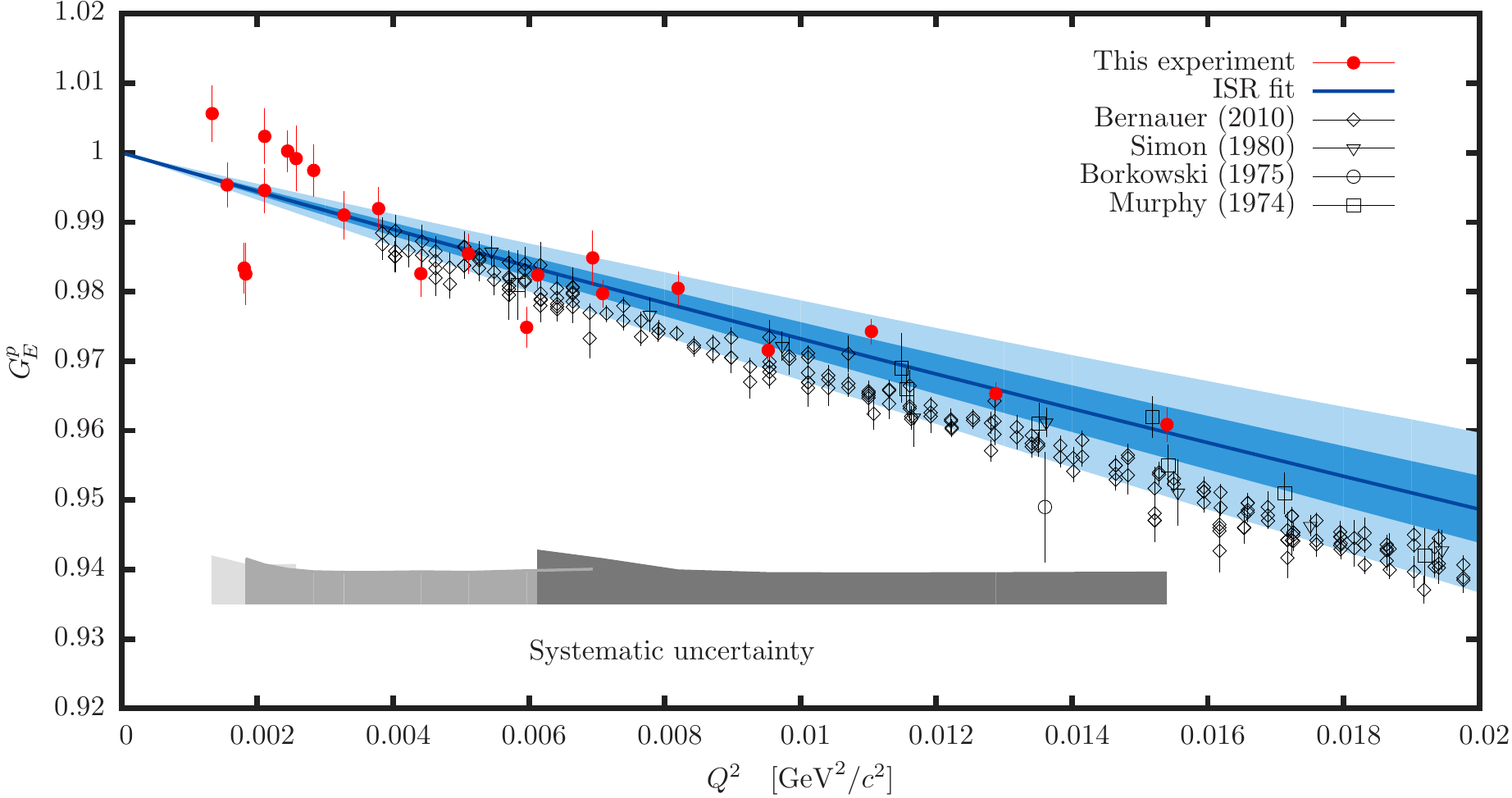}
 \caption{ (Color on-line) The proton electric form factor as a function of $Q^2(=Q^2_{\mathrm{Out}})$. 
 Empty black points show 
 previous data~\cite{Bernauer2010, SimonMAINZ,Murphy, BorkowskiMAINZ}. 
The results of this experiment are shown with full red circles.  The error bars show 
statistical uncertainties. Gray structures at the bottom shows the systematic uncertainties 
for the three energy settings. The curve corresponds to a polynomial 
fit to the  data defined by Eq.~(\ref{eq2}). The inner and the outer bands around the fit show its uncertainties, caused by the statistical and systematic uncertainties of the data, respectively.
 \label{fig_FF}}
\end{figure*}

The remaining inconsistencies between the data and simulation could be due to the 
higher-order effects that are missing in the simulation or unresolved backgrounds.
However, they could also be attributed to the difference between the true values of $G_E^{p}$ 
and the model used in the simulation. Hence, the results presented in Fig.~\ref{fig_Results} 
may also be considered in reverse. Assuming that the theoretical description of radiative 
corrections is flawless and that background processes are well 
under control, the differences between data and simulation have been used to extract new 
values of the proton charge form factor. 
We have determined $G_E^{p}$ for $0.001 \leq  Q^2 \leq 0.017\,\mathrm{GeV}^2/c^2$,
thus significantly extending the low $Q^2$-range of available data.  
The new values shown in Fig.~\ref{fig_FF} are consistent with 
results of previous measurements~\cite{SimonMAINZ, Murphy, BorkowskiMAINZ, Bernauer2010}
in the region of overlap. 
The extracted new $G_E^p$ values were compared to the polynomial
\begin{equation}
G(Q^2)= 1 - \frac{r_p^2\,Q^2}{6\,\hbar^2}  + \frac{a\,Q^4}{120\,\hbar^4} - 
   \frac{b\,Q^6}{5040\,\hbar^6}\,,~~\label{eq2}
\end{equation}
where parameters $a=(2.59\pm0.194)\,\mathrm{fm}^4$ and $b=(29.8\pm14.71)\,\mathrm{fm}^6$, 
which determine the curvature of the fit, were taken from Ref.~\cite{distler}.
The three data sets were fit with a common parameter for the radius, $r_p$, 
but with different renormalisation factors, $n_{E_0}$, for each energy.
In terms of this fit with $18$ degrees of freedom  and $\chi^2$ of $58.0$, 
the normalisations and the radius were determined to be 
$n_{195}=1.001 \pm 0.002_{\mathrm{stat}} \pm 0.003_{\mathrm{syst}}$,
$n_{330}=1.002 \pm 0.001_{\mathrm{stat}} \pm 0.003_{\mathrm{syst}}$,
$n_{495}=1.005 \pm 0.003_{\mathrm{stat}} \pm 0.007_{\mathrm{syst}}$,
and $r_p=(0.810 \pm 0.035_{\mathrm{stat}} \pm 0.074_{\mathrm{syst}} \pm 0.003_{\Delta a, \Delta b})\,\mathrm{fm}$. The reduced $\chi^2$ of $3.2$ 
per degree of freedom (statistical uncertainties only) indicates that the results are 
dominated by systematic effects. Due to the limiting backgrounds 
and corresponding systematic uncertainties, we are 
unable to distinguish convincingly  between the CODATA and 
the muonic hydrogen radii. However, we have proven the technique 
of initial state radiation to be a viable method for investigating the 
electromagnetic structure of the nucleon at 
extremely small $Q^2$. This has motivated further experiments of its kind. 
Utilising a gaseous point-like jet target together with a redesigned 
spectrometer entrance flange will significantly reduce instrumental backgrounds in
the planned followup experiment~\cite{JetISR}  thereby extending $G_E^p$ down 
to $Q^2\approx 2\cdot10^{-4}\,\mathrm{GeV}^2/c^2$. 



\begin{acknowledgments}
The authors would like to thank the MAMI accelerator group for the excellent beam
quality which made this experiment possible. This work is supported by the 
Federal State of Rhineland-Palatinate, by the Deutsche Forschungsgemeinschaft 
with the Collaborative Research Center 1044, by the Slovenian Research 
Agency under Grant Z1-7305 and U.~S. Department of Energy under Award Number 
DE-FG02-96ER41003.
\end{acknowledgments}

\bibliography{ISRPaper}

\begin{thebibliography}{24}%
\makeatletter
\providecommand \@ifxundefined [1]{%
 \@ifx{#1\undefined}
}%
\providecommand \@ifnum [1]{%
 \ifnum #1\expandafter \@firstoftwo
 \else \expandafter \@secondoftwo
 \fi
}%
\providecommand \@ifx [1]{%
 \ifx #1\expandafter \@firstoftwo
 \else \expandafter \@secondoftwo
 \fi
}%
\providecommand \natexlab [1]{#1}%
\providecommand \enquote  [1]{``#1''}%
\providecommand \bibnamefont  [1]{#1}%
\providecommand \bibfnamefont [1]{#1}%
\providecommand \citenamefont [1]{#1}%
\providecommand \href@noop [0]{\@secondoftwo}%
\providecommand \href [0]{\begingroup \@sanitize@url \@href}%
\providecommand \@href[1]{\@@startlink{#1}\@@href}%
\providecommand \@@href[1]{\endgroup#1\@@endlink}%
\providecommand \@sanitize@url [0]{\catcode `\\12\catcode `\$12\catcode
  `\&12\catcode `\#12\catcode `\^12\catcode `\_12\catcode `\%12\relax}%
\providecommand \@@startlink[1]{}%
\providecommand \@@endlink[0]{}%
\providecommand \url  [0]{\begingroup\@sanitize@url \@url }%
\providecommand \@url [1]{\endgroup\@href {#1}{\urlprefix }}%
\providecommand \urlprefix  [0]{URL }%
\providecommand \Eprint [0]{\href }%
\@ifxundefined \urlstyle {%
  \providecommand \doi  [0]{\begingroup \@sanitize@url \@doi}%
  \providecommand \@doi [1]{\endgroup \@@startlink {\doibase
  #1}doi:\discretionary {}{}{}#1\@@endlink }%
}{%
  \providecommand \doi  [0]{doi:\discretionary{}{}{}\begingroup
  \urlstyle{rm}\Url }%
}%
\providecommand \doibase [0]{http://dx.doi.org/}%
\providecommand \Doi [0]{\begingroup \@sanitize@url \@Doi }%
\providecommand \@Doi  [1]{\endgroup\@@startlink{\doibase#1}\@@Doi}%
\providecommand \@@Doi [1]{#1\@@endlink}%
\providecommand \selectlanguage [0]{\@gobble}%
\providecommand \bibinfo  [0]{\@secondoftwo}%
\providecommand \bibfield  [0]{\@secondoftwo}%
\providecommand \translation [1]{[#1]}%
\providecommand \BibitemOpen [0]{}%
\providecommand \bibitemStop [0]{}%
\providecommand \bibitemNoStop [0]{.\EOS\space}%
\providecommand \EOS [0]{\spacefactor3000\relax}%
\providecommand \BibitemShut  [1]{\csname bibitem#1\endcsname}%
\bibitem [{\citenamefont {Mohr}\ \emph {et~al.}(2016)\citenamefont {Mohr},
  \citenamefont {Newell},\ and\ \citenamefont {Taylor}}]{CODATA2014}%
  \BibitemOpen
  \bibfield  {author} {\bibinfo {author} {\bibfnamefont {P.~J.}\ \bibnamefont
  {Mohr}}, \bibinfo {author} {\bibfnamefont {D.~B.}\ \bibnamefont {Newell}}, \
  and\ \bibinfo {author} {\bibfnamefont {B.~N.}\ \bibnamefont {Taylor}},\ }\Doi
  {10.1103/RevModPhys.88.035009} {\bibfield  {journal} {\bibinfo  {journal}
  {Rev. Mod. Phys.},\ }\textbf {\bibinfo {volume} {88}},\ \bibinfo {pages}
  {035009} (\bibinfo {year} {2016})}\BibitemShut {NoStop}%
\bibitem [{\citenamefont {Pohl}\ \emph {et~al.}(2010)\citenamefont {Pohl} \emph
  {et~al.}}]{Pohl2010}%
  \BibitemOpen
  \bibfield  {author} {\bibinfo {author} {\bibfnamefont {R.}~\bibnamefont
  {Pohl}} \emph {et~al.},\ }\href {http://dx.doi.org/10.1038/nature09250}
  {\bibfield  {journal} {\bibinfo  {journal} {Nature},\ }\textbf {\bibinfo
  {volume} {466}},\ \bibinfo {pages} {213} (\bibinfo {year}
  {2010})}\BibitemShut {NoStop}%
\bibitem [{\citenamefont {Antognini}\ \emph {et~al.}(2013)\citenamefont
  {Antognini} \emph {et~al.}}]{Antognini2013}%
  \BibitemOpen
  \bibfield  {author} {\bibinfo {author} {\bibfnamefont {A.}~\bibnamefont
  {Antognini}} \emph {et~al.},\ }\Doi {10.1126/science.1230016} {\bibfield
  {journal} {\bibinfo  {journal} {Science},\ }\textbf {\bibinfo {volume}
  {339}},\ \bibinfo {pages} {417} (\bibinfo {year} {2013})}\BibitemShut
  {NoStop}%
\bibitem [{\citenamefont {Pohl}\ \emph {et~al.}(2013)\citenamefont {Pohl},
  \citenamefont {Gilman}, \citenamefont {Miller},\ and\ \citenamefont
  {Pachucki}}]{PohlAnn}%
  \BibitemOpen
  \bibfield  {author} {\bibinfo {author} {\bibfnamefont {R.}~\bibnamefont
  {Pohl}}, \bibinfo {author} {\bibfnamefont {R.}~\bibnamefont {Gilman}},
  \bibinfo {author} {\bibfnamefont {G.~A.}\ \bibnamefont {Miller}}, \ and\
  \bibinfo {author} {\bibfnamefont {K.}~\bibnamefont {Pachucki}},\ }\Doi
  {10.1146/annurev-nucl-102212-170627} {\bibfield  {journal} {\bibinfo
  {journal} {Ann. Rev. Nucl. Part. Sci.},\ }\textbf {\bibinfo {volume} {63}},\
  \bibinfo {pages} {175} (\bibinfo {year} {2013})}\BibitemShut {NoStop}%
\bibitem [{\citenamefont {Carlson}(2015)}]{Carlson201559}%
  \BibitemOpen
  \bibfield  {author} {\bibinfo {author} {\bibfnamefont {C.~E.}\ \bibnamefont
  {Carlson}},\ }\Doi {http://dx.doi.org/10.1016/j.ppnp.2015.01.002} {\bibfield
  {journal} {\bibinfo  {journal} {Prog. Part. Nucl. Phys.},\ }\textbf {\bibinfo
  {volume} {82}},\ \bibinfo {pages} {59 } (\bibinfo {year} {2015})}\BibitemShut
  {NoStop}%
\bibitem [{\citenamefont {Arbuzov}\ \emph {et~al.}(1998)\citenamefont
  {Arbuzov}, \citenamefont {Kuraev}, \citenamefont {Merenkov},\ and\
  \citenamefont {Trentadue}}]{Arbuzov1998}%
  \BibitemOpen
  \bibfield  {author} {\bibinfo {author} {\bibfnamefont {A.~B.}\ \bibnamefont
  {Arbuzov}}, \bibinfo {author} {\bibfnamefont {E.~A.}\ \bibnamefont {Kuraev}},
  \bibinfo {author} {\bibfnamefont {N.~P.}\ \bibnamefont {Merenkov}}, \ and\
  \bibinfo {author} {\bibfnamefont {L.}~\bibnamefont {Trentadue}},\ }\href
  {http://stacks.iop.org/1126-6708/1998/i=12/a=009} {\bibfield  {journal}
  {\bibinfo  {journal} {JHEP},\ }\textbf {\bibinfo {volume} {1998}},\ \bibinfo
  {pages} {009} (\bibinfo {year} {1998})}\BibitemShut {NoStop}%
\bibitem [{\citenamefont {Aubert}\ \emph {et~al.}(2004)\citenamefont {Aubert}
  \emph {et~al.}}]{BabarJPSi}%
  \BibitemOpen
  \bibfield  {author} {\bibinfo {author} {\bibfnamefont {B.}~\bibnamefont
  {Aubert}} \emph {et~al.} (\bibinfo {collaboration} {BABAR Collaboration}),\
  }\Doi {10.1103/PhysRevD.69.011103} {\bibfield  {journal} {\bibinfo  {journal}
  {Phys. Rev. D},\ }\textbf {\bibinfo {volume} {69}},\ \bibinfo {pages}
  {011103} (\bibinfo {year} {2004})}\BibitemShut {NoStop}%
\bibitem [{\citenamefont {Vanderhaeghen}\ \emph {et~al.}(2000)\citenamefont
  {Vanderhaeghen}, \citenamefont {Friedrich}, \citenamefont {Lhuillier},
  \citenamefont {Marchand}, \citenamefont {Van~Hoorebeke},\ and\ \citenamefont
  {Van~de Wiele}}]{Vanderhaeghen2000}%
  \BibitemOpen
  \bibfield  {author} {\bibinfo {author} {\bibfnamefont {M.}~\bibnamefont
  {Vanderhaeghen}}, \bibinfo {author} {\bibfnamefont {J.~M.}\ \bibnamefont
  {Friedrich}}, \bibinfo {author} {\bibfnamefont {D.}~\bibnamefont
  {Lhuillier}}, \bibinfo {author} {\bibfnamefont {D.}~\bibnamefont {Marchand}},
  \bibinfo {author} {\bibfnamefont {L.}~\bibnamefont {Van~Hoorebeke}}, \ and\
  \bibinfo {author} {\bibfnamefont {J.}~\bibnamefont {Van~de Wiele}},\ }\Doi
  {10.1103/PhysRevC.62.025501} {\bibfield  {journal} {\bibinfo  {journal}
  {Phys. Rev. C},\ }\textbf {\bibinfo {volume} {62}},\ \bibinfo {pages}
  {025501} (\bibinfo {year} {2000})}\BibitemShut {NoStop}%
\bibitem [{\citenamefont {Maximon}\ and\ \citenamefont
  {Tjon}(2000)}]{Maximon2000}%
  \BibitemOpen
  \bibfield  {author} {\bibinfo {author} {\bibfnamefont {L.~C.}\ \bibnamefont
  {Maximon}}\ and\ \bibinfo {author} {\bibfnamefont {J.~A.}\ \bibnamefont
  {Tjon}},\ }\Doi {10.1103/PhysRevC.62.054320} {\bibfield  {journal} {\bibinfo
  {journal} {Phys. Rev. C},\ }\textbf {\bibinfo {volume} {62}},\ \bibinfo
  {pages} {054320} (\bibinfo {year} {2000})}\BibitemShut {NoStop}%
\bibitem [{\citenamefont {Arenh{\"o}vel}\ and\ \citenamefont
  {Drechsel}(1974)}]{ARENHOVEL}%
  \BibitemOpen
  \bibfield  {author} {\bibinfo {author} {\bibfnamefont {H.}~\bibnamefont
  {Arenh{\"o}vel}}\ and\ \bibinfo {author} {\bibfnamefont {D.}~\bibnamefont
  {Drechsel}},\ }\Doi {http://dx.doi.org/10.1016/0375-9474(74)90248-6}
  {\bibfield  {journal} {\bibinfo  {journal} {Nucl. Phys. A},\ }\textbf
  {\bibinfo {volume} {233}},\ \bibinfo {pages} {153 } (\bibinfo {year}
  {1974})},\ ISSN \bibinfo {issn} {0375-9474}\BibitemShut {NoStop}%
\bibitem [{\citenamefont {Roche}\ \emph {et~al.}(2000)\citenamefont {Roche}
  \emph {et~al.}}]{Roche2000}%
  \BibitemOpen
  \bibfield  {author} {\bibinfo {author} {\bibfnamefont {J.}~\bibnamefont
  {Roche}} \emph {et~al.},\ }\Doi {10.1103/PhysRevLett.85.708} {\bibfield
  {journal} {\bibinfo  {journal} {Phys. Rev. Lett.},\ }\textbf {\bibinfo
  {volume} {85}},\ \bibinfo {pages} {708} (\bibinfo {year} {2000})}\BibitemShut
  {NoStop}%
\bibitem [{\citenamefont {Mo}\ and\ \citenamefont {Tsai}(1969)}]{MoTsai}%
  \BibitemOpen
  \bibfield  {author} {\bibinfo {author} {\bibfnamefont {L.~W.}\ \bibnamefont
  {Mo}}\ and\ \bibinfo {author} {\bibfnamefont {Y.~S.}\ \bibnamefont {Tsai}},\
  }\Doi {10.1103/RevModPhys.41.205} {\bibfield  {journal} {\bibinfo  {journal}
  {Rev. Mod. Phys.},\ }\textbf {\bibinfo {volume} {41}},\ \bibinfo {pages}
  {205} (\bibinfo {year} {1969})}\BibitemShut {NoStop}%
\bibitem [{\citenamefont {Tsai}(1961)}]{Tsai}%
  \BibitemOpen
  \bibfield  {author} {\bibinfo {author} {\bibfnamefont {Y.-S.}\ \bibnamefont
  {Tsai}},\ }\Doi {10.1103/PhysRev.122.1898} {\bibfield  {journal} {\bibinfo
  {journal} {Phys. Rev.},\ }\textbf {\bibinfo {volume} {122}},\ \bibinfo
  {pages} {1898} (\bibinfo {year} {1961})}\BibitemShut {NoStop}%
\bibitem [{\citenamefont {Blomqvist}\ \emph {et~al.}(1998)\citenamefont
  {Blomqvist} \emph {et~al.}}]{Blomqvist}%
  \BibitemOpen
  \bibfield  {author} {\bibinfo {author} {\bibfnamefont {K.}~\bibnamefont
  {Blomqvist}} \emph {et~al.},\ }\Doi
  {http://dx.doi.org/10.1016/S0168-9002(97)01133-9} {\bibfield  {journal}
  {\bibinfo  {journal} {Nucl.~Instr.~and~Meth.~A},\ }\textbf {\bibinfo {volume}
  {403}},\ \bibinfo {pages} {263 } (\bibinfo {year} {1998})},\ ISSN \bibinfo
  {issn} {0168-9002}\BibitemShut {NoStop}%
\bibitem [{\citenamefont {Mihovilovi\v{c}}\ \emph {et~al.}(2014)\citenamefont
  {Mihovilovi\v{c}} \emph {et~al.}}]{MihovilovicMessina}%
  \BibitemOpen
  \bibfield  {author} {\bibinfo {author} {\bibfnamefont {M.}~\bibnamefont
  {Mihovilovi\v{c}}} \emph {et~al.},\ }\Doi {10.1051/epjconf/20147200017}
  {\bibfield  {journal} {\bibinfo  {journal} {EPJ Web Conf.},\ }\textbf
  {\bibinfo {volume} {72}},\ \bibinfo {pages} {00017} (\bibinfo {year}
  {2014})}\BibitemShut {NoStop}%
\bibitem [{\citenamefont {Bernauer}\ \emph {et~al.}(2014)\citenamefont
  {Bernauer} \emph {et~al.}}]{Bernauer2014}%
  \BibitemOpen
  \bibfield  {author} {\bibinfo {author} {\bibfnamefont {J.~C.}\ \bibnamefont
  {Bernauer}} \emph {et~al.},\ }\Doi {10.1103/PhysRevC.90.015206} {\bibfield
  {journal} {\bibinfo  {journal} {Phys. Rev. C},\ }\textbf {\bibinfo {volume}
  {90}},\ \bibinfo {pages} {015206} (\bibinfo {year} {2014})}\BibitemShut
  {NoStop}%
\bibitem [{\citenamefont {Drechsel}\ \emph {et~al.}(2007)\citenamefont
  {Drechsel}, \citenamefont {Kamalov},\ and\ \citenamefont {Tiator}}]{MAID}%
  \BibitemOpen
  \bibfield  {author} {\bibinfo {author} {\bibfnamefont {D.}~\bibnamefont
  {Drechsel}}, \bibinfo {author} {\bibfnamefont {S.}~\bibnamefont {Kamalov}}, \
  and\ \bibinfo {author} {\bibfnamefont {L.}~\bibnamefont {Tiator}},\ }\Doi
  {10.1140/epja/i2007-10490-6} {\bibfield  {journal} {\bibinfo  {journal} {Eur.
  Phys. J. A},\ }\textbf {\bibinfo {volume} {34}},\ \bibinfo {pages} {69}
  (\bibinfo {year} {2007})},\ ISSN \bibinfo {issn} {1434-6001}\BibitemShut
  {NoStop}%
\bibitem [{\citenamefont {Janssens}\ \emph {et~al.}(2008)\citenamefont
  {Janssens} \emph {et~al.}}]{Janssens2008}%
  \BibitemOpen
  \bibfield  {author} {\bibinfo {author} {\bibfnamefont {P.}~\bibnamefont
  {Janssens}} \emph {et~al.},\ }\Doi {10.1140/epja/i2008-10609-3} {\bibfield
  {journal} {\bibinfo  {journal} {Eur. Phys. J. A},\ }\textbf {\bibinfo
  {volume} {37}},\ \bibinfo {pages} {1} (\bibinfo {year} {2008})},\ ISSN
  \bibinfo {issn} {1434-6001}\BibitemShut {NoStop}%
\bibitem [{\citenamefont {Bernauer}\ \emph {et~al.}(2010)\citenamefont
  {Bernauer} \emph {et~al.}}]{Bernauer2010}%
  \BibitemOpen
  \bibfield  {author} {\bibinfo {author} {\bibfnamefont {J.~C.}\ \bibnamefont
  {Bernauer}} \emph {et~al.},\ }\Doi {10.1103/PhysRevLett.105.242001}
  {\bibfield  {journal} {\bibinfo  {journal} {Phys. Rev. Lett.},\ }\textbf
  {\bibinfo {volume} {105}},\ \bibinfo {pages} {242001} (\bibinfo {year}
  {2010})}\BibitemShut {NoStop}%
\bibitem [{\citenamefont {Simon}\ \emph {et~al.}(1980)\citenamefont {Simon},
  \citenamefont {Schmitt}, \citenamefont {Borkowski},\ and\ \citenamefont
  {Walther}}]{SimonMAINZ}%
  \BibitemOpen
  \bibfield  {author} {\bibinfo {author} {\bibfnamefont {G.}~\bibnamefont
  {Simon}}, \bibinfo {author} {\bibfnamefont {C.}~\bibnamefont {Schmitt}},
  \bibinfo {author} {\bibfnamefont {F.}~\bibnamefont {Borkowski}}, \ and\
  \bibinfo {author} {\bibfnamefont {V.}~\bibnamefont {Walther}},\ }\Doi
  {http://dx.doi.org/10.1016/0375-9474(80)90104-9} {\bibfield  {journal}
  {\bibinfo  {journal} {Nucl. Phys. A},\ }\textbf {\bibinfo {volume} {333}},\
  \bibinfo {pages} {381 } (\bibinfo {year} {1980})},\ ISSN \bibinfo {issn}
  {0375-9474}\BibitemShut {NoStop}%
\bibitem [{\citenamefont {Murphy}\ \emph {et~al.}(1974)\citenamefont {Murphy},
  \citenamefont {Shin},\ and\ \citenamefont {Skopik}}]{Murphy}%
  \BibitemOpen
  \bibfield  {author} {\bibinfo {author} {\bibfnamefont {J.~J.}\ \bibnamefont
  {Murphy}}, \bibinfo {author} {\bibfnamefont {Y.~M.}\ \bibnamefont {Shin}}, \
  and\ \bibinfo {author} {\bibfnamefont {D.~M.}\ \bibnamefont {Skopik}},\ }\Doi
  {10.1103/PhysRevC.9.2125} {\bibfield  {journal} {\bibinfo  {journal} {Phys.
  Rev. C},\ }\textbf {\bibinfo {volume} {9}},\ \bibinfo {pages} {2125}
  (\bibinfo {year} {1974})}\BibitemShut {NoStop}%
\bibitem [{\citenamefont {Borkowski}\ \emph {et~al.}(1974)\citenamefont
  {Borkowski}, \citenamefont {Peuser}, \citenamefont {Simon}, \citenamefont
  {Walther},\ and\ \citenamefont {Wendling}}]{BorkowskiMAINZ}%
  \BibitemOpen
  \bibfield  {author} {\bibinfo {author} {\bibfnamefont {F.}~\bibnamefont
  {Borkowski}}, \bibinfo {author} {\bibfnamefont {P.}~\bibnamefont {Peuser}},
  \bibinfo {author} {\bibfnamefont {G.}~\bibnamefont {Simon}}, \bibinfo
  {author} {\bibfnamefont {V.}~\bibnamefont {Walther}}, \ and\ \bibinfo
  {author} {\bibfnamefont {R.}~\bibnamefont {Wendling}},\ }\Doi
  {http://dx.doi.org/10.1016/0375-9474(74)90392-3} {\bibfield  {journal}
  {\bibinfo  {journal} {Nucl. Phys. A},\ }\textbf {\bibinfo {volume} {222}},\
  \bibinfo {pages} {269 } (\bibinfo {year} {1974})},\ ISSN \bibinfo {issn}
  {0375-9474}\BibitemShut {NoStop}%
\bibitem [{\citenamefont {Distler}\ \emph {et~al.}(2011)\citenamefont
  {Distler}, \citenamefont {Bernauer},\ and\ \citenamefont
  {Walcher}}]{distler}%
  \BibitemOpen
  \bibfield  {author} {\bibinfo {author} {\bibfnamefont {M.~O.}\ \bibnamefont
  {Distler}}, \bibinfo {author} {\bibfnamefont {J.~C.}\ \bibnamefont
  {Bernauer}}, \ and\ \bibinfo {author} {\bibfnamefont {T.}~\bibnamefont
  {Walcher}},\ }\Doi {http://dx.doi.org/10.1016/j.physletb.2010.12.067}
  {\bibfield  {journal} {\bibinfo  {journal} {Phys. Lett. B},\ }\textbf
  {\bibinfo {volume} {696}},\ \bibinfo {pages} {343 } (\bibinfo {year}
  {2011})},\ ISSN \bibinfo {issn} {0370-2693}\BibitemShut {NoStop}%
\bibitem [{\citenamefont {Merkel~(spokesperson)}(2016)}]{JetISR}%
  \BibitemOpen
  \bibfield  {author} {\bibinfo {author} {\bibfnamefont {H.}~\bibnamefont
  {Merkel~(spokesperson)}},\ }\href@noop {} {\bibfield  {journal} {\bibinfo
  {journal} {MAMI proposal A1/02-16}} (\bibinfo {year} {2016})}\BibitemShut
  {NoStop}%
\end{thebibliography}%

\end{document}